\documentstyle[11pt,epsfig]{article}
\renewcommand{\title}[1]{{\Large\bf\mbox{}\\\medskip#1\bigskip\medskip\\}}
\renewcommand{\author}[1]{{\large #1\smallskip\\}}
\newcommand{\address}[1]{{\em #1\medskip\\}}
\def\be{\begin{equation}}
\def\ee{\end{equation}}
\def\bea{\begin{eqnarray}}
\def\eea{\end{eqnarray}}
\def\ba{\begin{array}}
\def\ea{\end{array}}
\def\a{\alpha}
\def\b{\beta}

\def\d{\delta}

\def\0{$\Gamma_0$}

\def\Mo{M\"obius }

\def\p{\phi}

\def\t{\theta}

\def\M{{\cal M}}
\def\N{{\cal N}}

\begin{document}
\begin{center}

\title{Close-packed dimers on nonorientable surfaces}
% New product identities}
  
\author{Wentao T. Lu and F. Y. Wu}
\address{Department of Physics\\
 Northeastern University, Boston,
Massachusetts 02115}

\begin{abstract}
The problem of enumerating dimers  on 
  an $\M \times \N$ net embedded on non-orientable surfaces is considered.
We solve
{\it both} the \Mo strip and Klein bottle problems for all $\M$ and $\N$
with the aid of imaginary dimer weights.  
The use of imaginary weights simplifies the analysis, and as a result we
obtain  new compact solutions
in the form of double products.  The compact expressions
also permit us to establish a general reciprocity theorem.
\end{abstract}
\end{center}

%\vskip 1cm
%\pacs{05.50.+q}

{\bf keywords}: close-packed dimers, non-orientable surfaces,
reciprocity theorem

%\maketitle

\vskip 1cm
%\pacs{05.50.+q}

\section{Introduction}
A seminal development in modern lattice statistics is the solution of 
enumerating close-packed dimers, or perfect matchings, on a finite $\M \times \N$ simple-quartic 
net obtained by
Kasteleyn \cite{kas} and by Temperley and Fisher \cite{temp, fisher61}
more than 40 years ago.
In their  solutions the  simple-quartic net is assumed to possess 
free or periodic boundary conditions \cite{kas}.
% Subsequently McCoy and Wu \cite{mccoy} solved the problem 
% for cylindrical boundary conditions. 
In view of the connection 
 with the conformal field theory \cite{henk},
where the boundary conditions play a crucial role,
there has been considerable renewed interest to
consider lattice models on non-orientable surfaces \cite{shrock,luwu99,tzeng,luwu01,Kaneda}.
 For close-packed dimers   the present authors \cite{luwu99} have
obtained the generating function
for 
% an $\M \times \N$ net embedded on 
the \Mo strip and 
the Klein bottle for  even $\M$ and $\N$.
 Independently, 
 Tesler \cite{tesler} has
solved the problem of  perfect matchings on \Mo strips
and deduced solutions 
 in terms of a $q$-analogue of the Fibonacci numbers 
for all $\{\M,\N\} $.
It turns out that
the explicit expression of  the  generating function
depends crucially on whether $\M$ and $\N$ being even or odd,
and the analysis differs considerably when  either $\M$ or $\N$ is odd.
In this paper  we consider the general $\{\M,\N\} $ 
problems  for {\it both}
 the \Mo strip  and the Klein bottle by introducing imaginary dimer weights.
The use of   imaginary weights  simplifies the analysis and as a result we
 obtain  compact expressions of the solutions without the recourse of Fibonacci numbers.
The compact expressions of the solutions also permit us to establish
a reciprocity theorem on the enumeration of dimers.
%Our analysis also leads to  new product identities that have not been previously noted.
 
\section{Summary of results}
For the convenience of references, we first summarize 
our main results.  
Details of  derivation  will be presented in subsequent sections.

\medskip
Consider an $\M \times \N$ simple-quartic net consisting
$\M\N$ sites arranged in an array of $\M$ rows and $\N$ columns.
 The net forms a \Mo strip if
there is a twisted boundary condition in the horizontal direction as shown in Fig. 1,
and a Klein bottle if, in addition to the twisted boundary condition, there 
is also a periodic boundary condition in the vertical direction.
    Let
the dimer weights be $z_h$ and $z_v$, respectively, in the 
horizontal and vertical directions.
We are interested in the close-form evaluation of the
dimer generating function
\be
Z_{{\cal M},{\cal N}}(z_h,z_v)=\sum z_h^{n_h}z_v^{n_v} \label{gen}
%=z_h^{{\cal MN}/2}\sum \tau^{n_v}
\ee
where 
$n_h, n_v$ are, respectively, the number of horizontal and vertical dimers, and
the summation is taken over all close-packed dimer coverings of the net.
% It is clear that the generating function vanishes identically unless $\M\N=$ even.

\medskip
Our results are as follows:
  For both ${\cal M}$ and ${\cal N}$ even, we have \cite{luwu99} 
\bea
Z_{\cal M,N}^{\rm Mob}(z_h,z_v)&=&\prod_{m=1}^{{\cal M}/2}\prod_{n=1}^{{\cal N}/2}
\Bigg[4z_h^2\sin^2 {(4n-1)\pi\over 2{\cal N}} +
4z_v^2\cos^2 {m\pi\over {\cal M}+1} \Bigg],\label{Mob-1}
\\
Z^{\rm Kln}_{\cal M,N}(z_h,z_v)&=&\prod_{m=1}^{{\cal M}/2}\prod_{n=1}^{{\cal N}/2}
\Bigg[4z_h^2\sin^2 {(4n-1)\pi\over 2{\cal N}}  +
4z_v^2\sin^2 {(2m-1)\pi\over {\cal M}} \Bigg], \label{Kln-1}
\eea
where the superscripts refer to the type of the nonorientable surface under
consideration.

\medskip
For ${\cal M}$ even and ${\cal N}$ odd, we have
\bea
Z_{\cal M,N}^{\rm Mob}(z_h,z_v)&=&{\rm Re}\Bigg[(1-i)
\prod_{m=1}^{{\cal M}/2}\prod_{n=1}^{\cal N}
\Bigg(2i(-1)^{{{\cal M}\over 2}+m+1}z_h\sin {(4n-1)\pi\over 2{\cal N}} 
+2z_v\cos {m\pi\over {\cal M}+1} \Bigg)\Bigg],\label{Mob-2}
\\
Z_{\cal M,N}^{\rm Kln}(z_h,z_v)&=&{\rm Re}\Bigg[(1-i)
\prod_{m=1}^{{\cal M}/2}\prod_{n=1}^{\cal N}
\Bigg(2i(-1)^{{{\cal M}\over 2}+m+1}z_h\sin {(4n-1)\pi\over 2{\cal N}} 
+2z_v\sin {(2m-1)\pi\over {\cal M}} \Bigg)\Bigg],\nonumber \\
\label{Kln-2}
\eea
 and for ${\cal M}$ odd and ${\cal N}$ even, we have
\bea
Z_{\cal M,N}^{\rm Mob}(z_h,z_v)
&=&z_h^{-{\cal N}/2}\prod_{n=1}^{{\cal N}/2}\prod_{m=1}^{({\cal M}+1)/2}
\Bigg[4z_h^2\sin^2 {(4n-1)\pi\over 2{\cal N}} 
+4z_v^2\cos^2 {m\pi\over {\cal M}+1} \Bigg],\label{Mob-3}
\\
Z_{\cal M,N}^{\rm Kln}(z_h,z_v)
&=&z_h^{-{\cal N}/2}\prod_{n=1}^{{\cal N}/2}\prod_{m=1}^{({\cal M}+1)/2}
\Bigg[4z_h^2\sin^2 {(4n-1)\pi\over 2{\cal N}} 
+4z_v^2\sin^2 {(2m-1)\pi\over {\cal M}} \Bigg]. \label{Kln-3}
\eea
 For ${\cal M}$ and ${\cal N}$ both odd, the generating function is zero.

%%%%%%%%%%%%%%%%%%%%%%%%%%%%%%%%%%%%%%%%%%%%%%%%%%%%%
\section{The \Mo strip}
%%%%%%%%%%%%%%%%%%%%%%%%%%%%%%%%%%%%%%%%%%%%%%%%%%%%%
 It is well-known that   there is a one-one correspondence between 
dimer coverings and terms in a Pfaffian defined by the dimer weights.
However, since terms in the Pfaffian generally do not possess the same sign,
the evaluation of the Pfaffian does not necessarily produce the 
desired dimer generating function.
 The crux of matter is to attach signs, or more generally factors, to the dimer
weights so that all terms in the Pfaffian  have the same sign, and the task
is reduced to that of evaluating a Pfaffian.

\medskip
These   tasks were first  achieved by Kasteleyn 
\cite{kas} for the simple-quartic  dimer lattice with
free  and periodic   boundary conditions,
 who showed how to attach signs to dimer weights
and how to evaluate the Pfaffian.
Soon after the publication of \cite{kas}, T. T. Wu \cite{wu}
pointed out that the structure of the Pfaffian, and hence its evaluation, can be simplified
 if a factor $i$ is associated 
to dimer weights in one spatial direction.
Indeed, the Wu prescription 
 requires only uniformly directed lattice edges with
 the association of a factor $i$
to dimer weights in the direction in which the number of lattice sites is odd.
(If the number of lattice sites is even in both directions, then the factor $i$ can
be associated to dimers in either direction.)
 For $\{\M,\N\} =\{ {\rm even, odd} \}$ for example,
one replaces $z_h$ by $iz_h$.

\medskip
To see that the Wu procedure is correct,
one considers a standard dimer covering $C_0$ in which the lattice is
covered only by  parallel dimers
with real weights.  
Then, the two terms in the Pfaffian corresponding to
$C_0$ and any other dimer covering $C_1$  have the same sign,
 since the superposition polygon produced by $C_0$ and $C_1$ 
contains an even number of arrows pointing in 
one direction as well as a factor $i^{4n+2} = -1$, where $n$ is a nonnegative
integer.  Namely, the superposition polygon is ``clockwise-odd"  \cite{wu}.
 This use of imaginary dimer weights is the starting point
of our analysis.

\subsection{${\cal M}=$ even \Mo strip}
%%%%%%%%%%%%%%%%%%%%%%%%%%%%%%%%%%%%%%%%%%%%%%%%%%%%%%%%
For $\M$ even and $\N=$ even or odd,
we write for definiteness  $\M = 2M$, $\N = N$ where $M,N$ are positive integers.
We orient lattice edges as shown in Fig. 1(a).
For the time being consider  more generally  that
the horizontal dimers connecting the first and $\N$-th columns have weights $z$
and the associated generating function 
   \be
Z^{\rm Mob}_{\cal M,N}(z_h,z_v;z)=\sum_{m=0}^{2M}z^mT_m \label{t}
\ee
 where $T_m\equiv T_m(z_h,z_v)$ is a multinomial in $z_h$ and $z_v$ with strictly
positive coefficients.   The desired result is then obtained by
setting $z=z_h$.   Note that $T_0$ is precisely the generating function with 
free boundary conditions.

\begin{figure}[htbp]
\center{\rule{5cm}{0.mm}}
\rule{5cm}{0.mm}
\vskip -0.2cm
%\hskip -2.0cm
\epsfig{figure=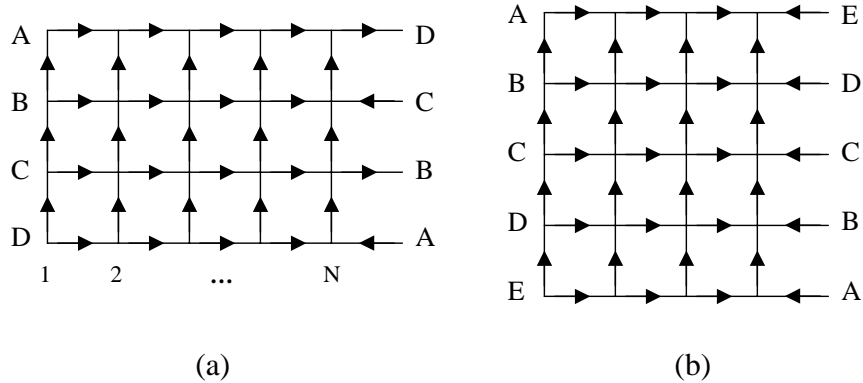,width=4.5in}
\vskip -0.2cm
\caption{Edge orientations of \Mo strips with twisted boundary conditions
in the horizontal direction.  A, B, C, D, E denote repeated sites.
(a). ${\cal M}=4$, ${\cal N}=5$.
(b). ${\cal M}=5$, ${\cal N}=4$.}
\label{dim-fig}
\end{figure}

\medskip
Attach a factor $i$ to all (horizontal) dimer weights $z_h$.   This 
 leads us to consider the
  antisymmetric matrix
\be
A(z)=iz_h(F_N-F_N^T)\otimes I_{2M}+z_vI_N\otimes(F_{2M}-F_{2M}^T)
+z(K_N+K_N^T)\otimes J_{2M} \label{matrixA}
\ee
where $I_N$ is the  $N\times N$ identity matrix, $F^T$ is the transpose of $F$, and
 $F_{2M}$, $K_N$ and $J_{2M}$ are  matrices of the order given by the subscripts,
\be
F_N = \pmatrix {0 & 1 & 0 &\cdots & 0 \cr
                  0 & 0  & 1 &\cdots & 0 \cr
                  \vdots & \vdots & \vdots & \ddots & \vdots \cr
                  0 & 0  & 0 &\cdots & 1 \cr
0 & 0  & 0 &\cdots & 0},\quad
K_N=\pmatrix{0&0&\cdots&0\cr \vdots&\vdots&\ddots&\vdots\cr
0&0&\cdots&0\cr
1&0&\cdots&0\cr}, \quad
 J_{2M}=\pmatrix{&&&&1\cr &&&-1&\cr
&&\cdots&&\cr
&1&&&\cr
-1&&&\cr}.
\ee
Now, the Pfaffian of $A(z)$ gives the correct generating function $T_0$
in the case of $z=0$ \cite{wu}.   For general $z$
we have the following result:

\medskip
{\it Theorem:
The dimer generating function for the simple-quartic net
with a twisted boundary condition in
the horizontal direction is }
\bea
Z^{\rm Mob}_{\cal M,N}(z_h,z_v;z_h) 
%&=& X_0 + z_hX_1 +  z_h^2X_2 +z_h^3X_3 \nonumber \\
  &=& {1\over 2}\Big[
(1-i){\rm Pf}{ A}(iz_h)+(1+i){\rm Pf}{ A}(-iz_h)\Big] \nonumber \\
&=& {\rm Re} \Big[(1-i){\rm Pf} A(iz_h)\Big].
 \label{Z-Pf-Mob}
\eea

\medskip
{\it Proof}:
It is clear that the term in (\ref{t}) corresponding to the configuration
$C_0$ ($m=0$) has the correct sign.  For any other dimer configuration $C_1$,
 the superposition of $C_0$ and $C_1$ forms superposition polygons
containing  $z$ edges.
We have the following  facts which can be readily verified:
 
(i) The sign of a superposition polygon remains unchanged under deformation of {\it its}
border which leaves $n_z$, the number of $z$ edges it contains, invariant.

(ii) Deformations of the border of a superposition polygon can change $n_z$ 
only by multiples of 2, and the sign of the superposition polygon reverses whenever
$n_z$ changes by 2.
 
(iii) Superposition polygons having 0 or 1 $z$ edges have the sign $+$.

\noindent
As a result, we obtain
\bea
{\rm Pf}A(z)& \equiv& \sqrt{|A(z)|} \nonumber \\
  &=& X_0+zX_1-z^2X_2-z^3X_3 \label{PfA(z)}
\eea
where $|\cdot |$ denotes the determinant of $\cdot$ and
\be
 X_\a=T_\a+z^4T_{\a+4}+z^8T_{\a+8}+\cdots, \hskip 1cm \a=0,1,2,3.
\ee
The theorem is now a consequence of the fact $Z^{\rm Mob}_{\cal M,N}(z_h,z_v;z_h) 
=X_0 + z_hX_1 +  z_h^2X_2 +z_h^3X_3$.

\medskip
Remarks:
The theorem holds also for the Klein bottle which, in addition to a
twisted boundary condition in the horizontal direction, has  a periodic
boundary condition in the vertical direction (see below).
%  This follows from the fact that (i) - (iii) remain valid.
For the M\"obius strip we have  $X_1=X_3=0$ when $N=$ even.

\medskip
 It now remains to evaluate PfA($\pm iz_h$).
To evaluate PfA($\pm iz_h)=\sqrt {|A(\pm i z_h)|}$ we make use of the fact  that,
 since $F_{2M}-F^T_{2M}$ commutes with $J_{2M}$,
the $ 2MN \times 2MN$ matrix 
$A(z)$ can be diagonalized in the
$2M$-subspace \cite{luwu99}. 
 Introducing  the $2M\times 2M$ matrix $U$  whose elements are
\bea
U_{m,m'} &=& i^m\sqrt{2\over {2M+1}}\> \>\sin 
\biggl( {{mm'\pi}\over {2M+1}}\biggr) \nonumber \\                
(U^{-1})_{m,m'}& =&  (-i)^{m'}\sqrt{2\over {2M+1}} \>\sin
 \biggl( {{mm'\pi}\over {2M+1}}\biggr),
 \hskip 1cm m,m'= 1,2,...,2M, \label{UMob}
\eea
we find
 \bea
 (U^{-1}(F_{2M} -F_{2M}^T)U)_{m,m'} &=& (2i\cos \phi_m)\ \d_{m,m'} \nonumber \\
(U^{-1}J_{2M} U)_{m,m'}&=& i\ (-1)^{M+m} \ \d_{m,m'},
\hskip 0.5cm
m,m'=1,2,...,2M, \label{eigen}
\eea
 where $\phi_m = {m\pi/ ({2M+1}})$.
Thus, we can replace the $2M\times 2M$ matrices in 
(\ref{matrixA}) 
 by their
respective eigenvalues,    and express $|A(z)|$ as a product of
the replaced determinants, namely, 
\be
| A(z)| =i^{2MN}\prod_{m=1}^{2M}| A_N^{(m)}(z)| \label{A(z)-N}
\ee
where we have taken out a common factor $i$ from each element of the $N\times N$ matrix 
\be
A_N^{(m)}(z)=2z_v\cos\phi_m I_N+z_h(F_N-F^T_N)+(-1)^{M+m}z( K_N+K_N^T).\label{Mob-A_N(z)}
\ee
 The matrix $A_N^{(m)}(z)$ can be evaluated for general $z$ in terms of a
$q$-analogue of
Fibonacci numbers,  but
for our purposes when $z=\pm iz_h$, the matrix can be diagonalized directly.

\medskip
 Define the $N\times N$ matrix
\be
T_N= F_N+i(-1)^{M+m}K_N = \pmatrix{  0 & 1 & 0 &\cdots & 0 \cr
                  0 & 0  & 1 &\cdots & 0 \cr
                  \vdots & \vdots & \vdots & \ddots & \vdots \cr
                  0 & 0  & 0 &\cdots & 1 \cr
i(-1)^{M+m} & 0  & 0 &\cdots & 0},
\ee
we can rewrite ({\ref{Mob-A_N(z)}) when $z=iz_h$ as
\be
A_N^{(m)}(iz_h)=2z_v\cos\p_m I_N+z_h(T_N-T_N^\dagger).
\ee
Now  $T_N$ and $T_N^\dagger$ commute so they can
be diagonalized simultaneously with respective
eigenvalues $e^{i\t_n}$ and $e^{-i\t_n}$, where 
\footnote{A similar expression of $\t_n$  given in
(17) of Ref. \cite{luwu99} contains a typo where 
%a minus sign is missing on the rhs
$M+m+1$ in the exponent  should read $M+m$. 
This does not alter the results of Ref. \cite{luwu99}.}
 %$
\be
\t_n=(-1)^{M+m+1}(4n-1)\pi/2N.
\ee
 Thus, we obtain
\be
|A_N^{(m)}(iz_h)|=\prod_{n=1}^N \Bigg[2z_v\cos{{m\pi}\over {2M+1}}
+2i(-1)^{M+m+1}z_h\sin{{(4n-1)\pi}\over {2N}}\Bigg],
 \ee
and as a result
\be
|A(iz_h)|=\prod_{m=1}^{M}\prod_{n=1}^N\Bigg[2z_v\cos{{m\pi}\over {2M+1}}
+2i(-1)^{M+m+1}z_h\sin{{(4n-1)\pi}\over {2N}}\Bigg]^2 \label{a2}
\ee
where we have made use of the fact that $\cos \p_{2M+1-m} = - \cos \p_m$,
$(-1)^{2M+1-m}=-(-1)^{m}$, and  $i^{2MN} = (-1)^{MN}$.
 We thus obtain after taking the square root of (\ref{a2})
 \be
{\rm Pf}A(iz_h)=\prod_{m=1}^M\prod_{n=1}^N
\Bigg[2z_v\cos{{m\pi}\over {2M+1}}+2i(-1)^{M+m+1}z_h\sin{(4n-1)\pi\over 2N}
\Bigg].  \label{Pfzh}
\ee
  The substitution of (\ref{Pfzh}) into  (\ref{Z-Pf-Mob})  now yields   (\ref{Mob-2}).
For $N=$ even the Pfaffian (\ref{Pfzh}) is real and  (\ref{Mob-2})
reduces to (\ref{Mob-1}).    There is no such simplification for $N=$ odd.

%%%%%%%%%%%%%%%%%%%%%%%%%%%%%%%%%%%%%%%%%%%%%%%%%%%%%%%%
\subsection{${\cal M}=$ odd \Mo strip}
%%%%%%%%%%%%%%%%%%%%%%%%%%%%%%%%%%%%%%%%%%%%%%%%%%%%%%%%
For  $\M=$ odd and $\N=$ even or odd, we write, for definiteness,
$\M= 2M+1$, $\N =N$.  Since the number of rows $\M$ is odd, we now
attach a factor $i$ to dimers in the vertical direction.
Again, we assign weights $z$ to 
horizontal dimers connecting the first and $N$-th columns
and consider the generating function
  \be
Z^{\rm Mob}_{\cal M,N}(z_h,z_v;z)=\sum_{m=0}^{2M+1}z^mT_m \label{t1}
\ee
defined similar to (\ref{t}).  It is readily verified that,
with  lattice edge orientations  shown in Fig. 1(b),
all terms in $T_0$ have the same sign.  
It follows that we can use the theorem of the 
 preceding subsection 
 where Pf($A$) is the Pfaffian of the antisymmetric matrix
 \be
A(z)=z_h(F_N-F_N^T)\otimes I_{2M+1}-iz_vI_N\otimes(F_{2M+1}-F_{2M+1}^T)
+zG_N\otimes H_{2M+1},\label{matrixA-2}
\ee
with $G_N=K_N-K_N^T$ and
\be
H_{2M+1}=\pmatrix{&&&1\cr
&&1&\cr
&\cdots&&\cr
1&&&\cr}.
\ee
 Apply to (\ref{matrixA-2}) the unitary transformation (\ref{UMob})
 (with $2M$ replaced by $2M+1$).  The transformation diagonalizes
$(F_{2M+1}-F_{2M+1}^T)$ 
as in (\ref{eigen}) and, in addition,  produces
 \be
U^{-1}H_{2M+1}U=(-1)^MJ_{2M+1}.
\ee
Thus,  we obtain
\bea
{\bar A}(z)&\equiv &(I_N\otimes U^{-1})A(z)(I_N\otimes U) \nonumber \\
 &=& \Big[z_h(F_{N}-F_{N}^T)+2z_v\cos{\bar \p_m}I_N\Big]\otimes I_{2M+1}
-zG_N\otimes J_{2M+1}
\eea
where ${\bar \p_m} =m\pi/(2M+2)$. 

\medskip
Writing 
\be
B_N^{(m)}\equiv z_h(F_{N}-F_{N}^T)+2z_v\cos{\bar \p_m} I_N,
\ee
then the  matrix ${\bar A}(z)$ assumes a form  shown below in
 the case of $2M+1=5$,
\be
{\bar A}(z)= \pmatrix{
B_N^{(1)}&&&&zG_N\cr
&B_N^{(2)}&&-zG_N&\cr
&&B_N^{(3)}+zG_N&&\cr
&-zG_N&&B_N^{(4)}&\cr
zG_N&&&&B_N^{(5)}\cr}.
\ee
Interchanging rows and columns,
  this matrix can be changed into a block-diagonal form having the same determinant,
\be
\pmatrix{
B_N^{(1)}&zG_N&&&\cr
zG_N&B_N^{(5)}&&&\cr
&&B_N^{(2)}&-zG_N&\cr
&&-zG_N&B_N^{(4)}&\cr
&&&&B_N^{(3)}+zG_N\cr}.
\ee
 
\medskip
For general $M$ we define the $2N\times 2N$ matrix
\be
A_{2N}^{(m)}(z)= \pmatrix{B_N^{(m)}&(-1)^{M+m+1}zG_N\cr
(-1)^{M+m+1}zG_N&B_N^{(2M+1-m)}\cr},
 % \nonumber \\
% &=&z_h(F_N-F_N^T)\otimes I_2
 % +2z_v\cos\p_m I_N\otimes \pmatrix{1&0\cr 0&-1\cr}
% +(-1)^{M+m+1}zG_N\otimes\pmatrix{0&1\cr 1&0\cr},
\ee
and use the result 
\be
|B^{(M+1)}_N+zG_N|=\frac 1 2 \big[1+(-1)^N\big]z_h^{N-2}
(z_h+z)^2 \label{odd}
\ee
to obtain
   \be
|A(z)|=|{\bar A}(z)|=\frac 1 2 \big[1+(-1)^N\big]z_h^{N-2}
(z_h+z)^2 \cdot \prod_{m=1}^M|A_{2N}^{(m)}(z)|.\label{az}
\ee
 It therefore remains to evaluate $|A_{2N}^{(m)}(z)|$.

\medskip
The matrix $A_{2N}^{(m)}(z)$ can  be  diagonalized for $z=\pm iz_h$.
To proceed, it is convenient to multiply from the right by a $2N\times 2N$ matrix
(whose determinant is $(-1)^N$) to obtain
 \bea
{\bar A}_N^{(m)}(iz_h)&\equiv&
A_{2N}^{(m)}(iz_h)\Bigg[I_N\otimes\pmatrix{1&0\cr 0&-1\cr}\Bigg]\nonumber \\
&=&2z_v\cos{\bar \p}_m I_{2N}+
z_h(F_N-F_N^T)\otimes \pmatrix{1&0\cr 0&-1\cr}
+i(-1)^{M+m+1}z_hG_N\otimes\pmatrix{0&-1\cr 1&0\cr} \nonumber \\
&=& 2z_v\cos{\bar \p}_m I_{2N}+z_h(Q_{2N}-Q^\dagger_{2N}),
 \label{barA}
\eea
 where 
\be
Q_{2N}=\pmatrix{F_N&i(-1)^{M+m}K_N\cr -i(-1)^{M+m}K_N&-F_N\cr}.
\ee
Now  $Q_{2N}$ commutes with $Q^\dagger_{2N}$ so they can be 
diagonalized simultaneously with the respective eigenvalues
$e^{i{\bar \t}_n}$ and $e^{-i{\bar \t}_n}$,  where
${\bar \t}_n = ( {2n-1})/ {2N}, n=1,2,\cdots 2N$.
% Let the eigenvector $\Psi$ be $\Psi^T=(\cdots,\psi_n,\cdots)$,
% we obtain from the eigenequation $Q_{2N}\Psi=\l\Psi$ 
% \be
% \psi_{n+1}=\l\psi_n,\quad 
% \psi_{N+1}=i(-1)^{M+m+1}\l\psi_N,\quad 
% \psi_{N+n+1}=-\l\psi_{N+n},\quad \psi_{1}=i(-1)^{M+m}\l\psi_{2N}.
% \ee
% Here $1\leq n<N$.
% From the above equations, the component of the eigenvector assumes the form
% \be
% \psi_{n}=\a^{n},\quad  \psi_{N+n}=i(-1)^{M+m+n}\a^{N+n},\quad 1\leq n\leq N
% \ee
% with $\l=\a$ and
% \be
% \a^{2N}=(-1)^{N+1}
% \ee
% Since $N$ is even, we get the eigenvalue of $R_{2N}$ as
% $\l=e^{i\p_n}$ with ${\bar \t}_n=(2n-1)/2N$.
Thus we obtain
 \be
|A_{2N}^{(m)}(iz_h)|
=(-1)^N|{\bar A}_{2N}^{(m)}(iz_h)|
=(-1)^N\prod_{n=1}^{N}(4z_h^2\sin^2{\bar \t}_n+4z_v^2\cos^2{\bar \p}_m),
\ee
and, taking the square root of (\ref{az}) with $z$ replaced by $iz_h$,
\be
{\rm Pf}A(iz_h)=\frac 1 2 \big[1+(-1)^N\big](1+i)z_h^{N/2}\cdot
\prod_{n=1}^{N/2}\prod_{m=1}^{M}\Big[4z_h^2\sin^2{(2n-1)\pi\over 2N}
+4z_v^2\cos^2{m\pi\over 2M+2}\Big]. \label{Klna}
\ee
Therefore the generating function  vanishes identically
 for $N=$ odd.  For $N=$ even we
substitute (\ref{Klna}) into (\ref{Z-Pf-Mob}) and make use of the fact that
the two sets 
  $\sin ^2(2n-1)\pi/2N$ and
$\sin ^2(4n-1)\pi/2N$, $n=1,2,\cdots, N/2$ are identical.
This leads to the result (\ref{Mob-3}).
%%%%%%%%%%%%%%%%%%%%%%%%%%%%%%%%%%%%%%%%%%%%%%%%%%%%%%%%
\section{The Klein bottle}
%%%%%%%%%%%%%%%%%%%%%%%%%%%%%%%%%%%%%%%%%%%%%%%%%%%%%%%
A Klein bottle is constructed by  inserting ${\cal N}$ extra vertical edges 
with weight $z_v$ to the  boundary of the \Mo strips of Figs. 1(a) and 1(b), so
that there is  a periodic boundary condition in the vertical direction.
 The extra edges are  oriented upward as the other vertical edges.
The consideration of the  Klein bottle
parallels to that of the \Mo strip.
Again, we need to consider the cases of even and odd $\M$ separately.
\subsection{$\M=$ even Klein bottle}
%%%%%%%%%%%%%%%%%%%%%%%%%%%%%%%%%%%%%%%%%%%%%%%%%%%%%%%
For a $2M\times N$  Klein bottle with  
horizontal edges connecting the first and $\N$-th row having weights $z$,
 we have as in (\ref{t}) the generating function
   \be
Z^{\rm Kln}_{\cal M,N}(z_h,z_v;z)=\sum_{m=0}^{2M}z^mT_m.  \label{k}
\ee
The dimer weights now generate
 the antisymmetric matrix
\be
A^{\rm Kln}(z)=A(z)-z_vI_N\otimes(
K_{2M}-K_{2M}^T)
\ee
where $A(z)$ is given by (\ref{matrixA}).  Following the same analysis
as in the case of the \Mo strip, since (i) - (iii) still hold 
we find the desired dimer generating function 
again given by the theorem (\ref{Z-Pf-Mob}) or, explicitly, 
\be
Z^{\rm Kln}_{\cal M,N}(z_h,z_v;z_h)={1\over 2}\Big[
(1-i){\rm Pf}{ A}^{\rm Kln}(iz_h)+(1+i){\rm Pf}{ A}^{\rm Kln}(-iz_h)\Big]. \label{k3}
\ee

To evaluate   ${\rm Pf}A^{\rm Kln}(z)$, we note that the $2M \times 2M$ matrices
$(F_{2M}-K_{2M}-F_{2M}^T+K_{2M}^T)$ and $J_{2M}$ commute,
and can be diagonalized simultaneously by 
the $2M\times 2M$ matrix $W$ whose elements are
\bea
W_{mm'}&=&{1\over \sqrt{4M}}\Big[
e^{i(2m-1)(2m'-1)\pi/4M}-i(-1)^{m+m'+M}e^{-i(2m-1)(2m'-1)\pi/4M}\Big],
\nonumber \\
(W^{-1})_{mm'}&=&{1\over \sqrt{4M}}\Big[
e^{-i(2m-1)(2m'-1)\pi/4M}+i(-1)^{m+m'+M}e^{i(2m-1)(2m'-1)\pi/4M}\Big].
\eea
We find
 \bea
 [W^{-1}(F_{2M} -K_{2M}-F_{2M}^T+K_{2M}^T)W]_{m,m'} &=& (2i\sin { \alpha}_m)\ \d_{m,m'} \nonumber \\
(W^{-1}J_{2M} W)_{m,m'}&=& i\ (-1)^{M+m} \ \d_{m,m'},
\label{Kln-eigen}
\eea                 
where $\alpha_m=(2m-1)\pi/2M$. Diagonalizing $A(z)$ in the $2M$-subspace, we obtain
\be
| A^{\rm Kln}(z)| =i^{2MN}\prod_{m=1}^{2M}|A_N^{(m)}(z)|
\ee
where
\be
A_N^{(m)}(z)=2z_v\sin\alpha_m I_N+z_h(F_N-F_N^T)+(-1)^{M+m}z(K_N+K_N^T).
\ee
This expression is the same as
  (\ref{Mob-A_N(z)}) for the \Mo strip, except with the substitution 
of $\cos \phi_m$ by $\sin\alpha_m$.
Thus,  following the same analysis, we obtain
 \be
{\rm Pf}A(iz_h)=\prod_{m=1}^M\prod_{n=1}^N
\Bigg[2z_v\sin{{(2m-1)\pi}\over {2M}}+2i(-1)^{M+m+1}z_h\sin{(4n-1)\pi\over 2N}
\Bigg]. \label{kln2}
\ee
The substitution of (\ref{kln2}) into 
(\ref{k3}) now gives the result  (\ref{Kln-2}).
For $N=$ even, the Pfaffian (\ref{kln2}) is real and (\ref{Kln-2})
reduces to (\ref{Kln-1}).

%%%%%%%%%%%%%%%%%%%%%%%%%%%%%%%%%%%%%%%%%%%%%%%%%%%%%%%
\subsection{${\cal M}=$ odd Klein bottle}
%%%%%%%%%%%%%%%%%%%%%%%%%%%%%%%%%%%%%%%%%%%%%%%%%%%%%%%
For a $(2M+1)\times N$  Klein bottle,   the inserted vertical edges have 
dimer weights  $iz_v$.  The consideration then parallels that of the preceeding sections.
Particularly, the desired dimer generating function is also given by
(\ref{k3}), but now with  
  \be
A^{\rm Kln}(z)=A(z)+iz_vI_N\otimes(
K_{2M+1}-K_{2M+1}^T).
\ee
To evaluate  ${\rm Pf}A^{\rm Kln}(z)$, one again applies 
in the $2M+1$ subspace the unitary transformation
which diagonalizes
  $F_{2M+1}-
K_{2M+1}-F_{2M+1}^T+K_{2M+1}^T$.
Define
\bea
V_{mm'}&=&{1\over \sqrt{2M+1}} e^{im(2m'-1)\pi/(2M+1)}, \nonumber \\
%\exp\>i{m(2m'-1)\pi\over 2M+1},\nonumber \\
(V^{-1})_{mm'}&=&{1\over \sqrt{2M+1}} e^{-im'(2m-1)\pi/(2M+1)},
%\exp\>-i{m'(2m-1)\pi\over 2M+1},
\quad m,m'=1,2,...,2M+1.
\label{UKln}
\eea
Using the result 
\bea
[V^{-1}(F_{2M+1} -K_{2M+1}-F_{2M+1}^T+K_{2M+1}^T)V]_{m,m'} &=& (2i\sin { {\bar\alpha}}_m)\ \d_{m,m'} \nonumber \\
(V^{-1}H_{2M+1}V)_{mm'}&=& -e^{-i{\bar \a}_m}\d_{m,2M+2-m'},
\eea
where ${\bar \a} _m = (2m-1)\pi/(2M+1), m=1,2,\cdots,2M+1$,
then the matrix ${\bar A}(z)=(I_N\otimes V^{-1})A(z)(I_N\otimes V)$
assumes the form in the case of $2M+1=5$,
\be
{\bar A}(z)=
\pmatrix{
B_N^{(1)}&&&&-ze^{-i{\bar \a}_1}G_N\cr
&B_N^{(2)}&&-ze^{-i{\bar \a}_2}G_N&\cr
&&B_N^{(3)}+zG_N&&\cr
&-ze^{i{\bar \a}_2}G_N&&B_N^{(4)}&\cr
-ze^{i{\bar \a}_1}G_N&&&&B_N^{(5)}\cr}.
\ee
Here $B_N^{(m)}=z_hQ_N+2z_v\sin{\bar \a}_m$.
Again, interchanging rows and columns, 
one  changes ${\bar A}(z)$ into the
block-diagonal form
\be
\pmatrix{
B_N^{(1)}&-ze^{-i{\bar \a}_1}G_N&&&\cr
-ze^{i{\bar \a}_1}G_N&B_N^{(5)}&&&\cr
&&B_N^{(2)}&-ze^{-i{\bar \a}_2}G_N&\cr
&&-ze^{i{\bar \a}_2}G_N&B_N^{(4)}&\cr
&&&&B_N^{(3)}+zG_N\cr}.
\ee
Explicitly, for general $M$, the $m$-th block is  
\bea
A_{2N}^{(m)}(z)&=&\pmatrix{B_N^{(m)}&-ze^{-i{\bar \a}_m}G_N\cr
-ze^{i{\bar \a}_m}G_N&B_N^{(2M+1-m)}\cr}\nonumber \\
&=&z_h(F_N-F_N^T)\otimes I_2
+2z_v\sin{\bar \a}_m I_N\otimes \pmatrix{1&0\cr 0&-1\cr}
-zG_N\otimes\pmatrix{0&e^{-i{\bar \a}_m}\cr e^{i{\bar \a}_m}&0\cr} \nonumber \\
\eea
We proceed as in (\ref{barA}) by multiplying a $2N\times 2N$ matrix
whose determinant is $(-1)^N$ from the right, and obtain
\bea
{\bar A}_{2N}^{(m)}(iz_h)& \equiv &
A_{2N}^{(m)}(iz_h)\Bigg[I_N\otimes\pmatrix{1&0\cr 0&-1\cr}\Bigg]\nonumber \\
&=&2z_v\sin{\bar \a}_m I_{2N}+
z_h(F_N-F_N^T)\otimes \pmatrix{1&0\cr 0&-1\cr}
+iz_hG_N\otimes\pmatrix{0&e^{-i{\bar \a}_m}\cr -e^{i{\bar \a}_m}&0\cr} \nonumber \\
&=&2z_v\sin{\bar \a}_m I_{2N}+z_h(Q_{2N}-Q^\dagger_{2N}), \label{kln3}
 \eea
where
\be
Q_{2N}=\pmatrix{F_N&ie^{-i{\bar \a}_m}K_N\cr -ie^{i{\bar \a}_m}K_N&-F_N\cr}.
\ee
 Now $Q_{2N}$ commutes with $Q^\dagger_{2N}$, and they can be 
diagonalized simultaneously
% Let the eigenvector $\Psi$ be $\Psi^T=(\cdots,\psi_n,\cdots)$,
% we obtain from the eigenequation $Q_{2N}\Psi=\l\Psi$ 
% \be
% \psi_{n+1}=\l\psi_n,\quad 
% \psi_{N+1}=-ie^{-i\p_m}\l\psi_N,\quad 
% \psi_{N+n+1}=-\l\psi_{N+n},\quad \psi_{1}=ie^{i\p_m}\l\psi_{2N}.
% \ee
% Here $1\leq n<N$.
% From the above equations, the component of the eigenvector assumes
% the form
% \be
% \psi_{n}={\bar \a}^{n},\quad  \psi_{N+n}=i(-1)^{n}e^{-i\p_m}{\bar \a}^{N+n},
% \quad 1\leq n\leq N
% \ee
% with $\l={\bar \a}$ and
% \be
% {\bar \a}^{2N}=(-1)^{N+1}
% \ee
with the respective eigenvalues $e^{i{\bar \b}_n}$ and $e^{-i{\bar \b}_n}$
where ${\bar \b}_n=(2n-1)/2N , n=1,2,\cdots, 2N$.
% Since $N$ is even, we get the eigenvalue of $Q_{2N}$ as
% $\l=e^{i\t_n}$ with $\t_n=(2n-1)/2N$.
Thus we obtain
\be
\Big|{\bar A}_{2N}^{(m)}(iz_h)\Big|=\prod_{n=1}^{N}
\Bigg[4z_h^2\sin^2{(2n-1)\pi\over 2N}
+4z_v^2\sin^2{(2m-1)\pi\over {2M+1}}\Bigg]
\ee
Using (\ref{kln3}) and  (\ref{odd}), we get
\be
{\rm Pf}A(iz_h)={1\over 2}\Big(1+(-1)^N\Big)(1+i)z_h^{N/2}
\prod_{n=1}^{N/2}\prod_{m=1}^{M}\Bigg[4z_h^2\sin^2{(2n-1)\pi\over 2N}
+4z_v^2\sin^2{(2m-1)\pi\over {2M+1}}\Bigg].\label{k5}
\ee
Thus, the  generating function (\ref{k3}) vanishes identically
for  $N=$ odd.  For $N=$ even, we replace as before
$\sin^2 (2n-1)\pi/N$  by $\sin ^2(4n-1)\pi/N$.  The substitution of
 (\ref{k5}) into (\ref{k3}) now
  leads to the result (\ref{Kln-3}). 
%%%%%%%%%%%%%%%%%%%%%%%%%%%%%%%%%%%%%%%%
\section{A reciprocity theorem}
Using the explicit expression of dimer
 enumerations  on  a simple-quartic lattice with free boundaries,
 Stanley \cite{stanley}  has shown that the enumeration expression satisfies
a certain reciprocity relation, a relation  rederived recently
by  Propp \cite{propp01} from a combinatorial approach.
  Here, we show that our solutions of  dimer
enumerations lead to an extension of the reciprocity relation to 
 enumerations on cylindrical, toroidal, and
nonorientable surfaces.  

\medskip
We first consider  solutions (\ref{Mob-1}), (\ref{Kln-1}), 
(\ref{Mob-3}) and (\ref{Kln-3}) 
for $\N=$ even.  Writing
\be
T^{\rm Mob}({\cal M,N})=Z_{\cal M,N}^{\rm Mob}(1,1),\quad
T^{\rm Kln}({\cal M,N})=Z_{\cal M,N}^{\rm Kln}(1,1)
\ee
and using the identity \cite{gr}
\be
\prod_{k=0}^{n-1} \Bigg[x^2-2x \cos \Bigg(\a+\frac {2k\pi}{n}\Bigg) +1\Bigg] = 
x^{2n}-2x^n \cos( n\a) +1
%\prod_{n=1}^{N} \Bigg(x^2+2x \cos \frac {2n\pi}{2N+1} +1\Bigg) &=& \frac {x^{2N+1}+1}{x+1}
\ee
repeatedly,
we can rewrite our  solutions (for general $z_h$ and $z_v$)
 in the form of 
\bea
Z_{\cal M,N}^{\rm Mob}(z_h,z_v)
&=&z_h^{{\cal MN}/2}\prod_{m=1}^{[({\cal M}+1)/2]}(x_m^{{\cal N}}+x_m^{-{\cal N}})
\nonumber \\
&=&\sqrt{3-(-1)^{\cal M}\over 2}z_v^{{\cal MN}/2}\prod_{n=1}^{{\cal N}/2}
\Bigg[{y_n^{{\cal M}+1}+(-1)^{\cal M}y_n^{-{\cal M}-1}\over y_n+y_n^{-1}}\Bigg] \label{fib1}  \\
Z_{\cal M,N}^{\rm Kln}(z_h,z_v)
&=&z_h^{{\cal MN}/2}\prod_{m=1}^{[({\cal M}+1)/2]}(t_m^{{\cal N}}+t_m^{-{\cal N}})
\nonumber \\
&=&\sqrt{3-(-1)^{\cal M}\over 2}z_v^{{\cal MN}/2}\prod_{n=1}^{{\cal N}/2}
\Big[y_n^{\cal M}+(-y_n)^{-{\cal M}}+1+(-1)^{\cal M}\Big],\label{fib2}
\eea
where
\be
x_m=G\Bigg(\frac {z_v}{z_h}\cos \frac {m\pi} {\M+1}\Bigg) ,
\
y_n=G\Bigg(\frac {z_h}{z_v} \sin  \frac{(2n-1)\pi}  {2\N}\Bigg) ,
\
t_m=G \Bigg(\frac {z_v}{z_h} \sin\frac {(2m-1)\pi} {\M}\Bigg)
\ee
and $G(y)= y+\sqrt{y^2+1}  $.
 Thus,  the following reciprocity relations are obtained by inspection:
\bea
T^{\rm Mob}({\cal M,N})&=& T^{\rm Mob}({\cal M},-{\cal N})
= \epsilon_{\cal NM}\ T^{\rm Mob}(-M-2,N)
 \nonumber \\
T^{\rm Kln}({\cal M,N})&=&T^{\rm Kln}({\cal M},-{\cal N}) = \ \epsilon_{\cal NM} 
\ T^{\rm Kln}(-\M,\N),
\eea
where
\bea
\epsilon_{\cal NM} &=& (-1)^\M, \hskip .7cm {\cal N} = 2 \>\> ( {\rm mod\>\>}4) \nonumber \\
                   &=& +1, \hskip 1.4cm {\rm otherwise}.
\eea
There are no reciprocity relations for $\N=$ odd.

\medskip
We have carried out similar analyses for dimer enumerations on a simple-quartic 
net embedded on a cylinder and a torus, using the solutions given in
 \cite{mccoywu,kas},  and have
discovered  universal rules of associating reciprocity relations to specific 
boundary conditions.
Generally, there are 3 different boundary conditions, or ``matchings",  
 that can be imposed between  2 opposite boundaries of an $\M\times\N$ net.  The
conditions can be  {\it twisted} such as
those shown in the horizontal direction in Fig. 1, {\it periodic} such as on a torus,
 or {\it free} which means free standing.  
%Thus, we have a \Mo strip if the 2 boundary conditions are free and twisted,
%%a Klein bottle if they are periodic and twisted, a cylinder if periodic and free,
%and a torus if periodic in both directions.
To establish the convention we shall
refer to the boundary condition  between  the first and the $\N$-th columns
as (the boundary condition) in the $\N$ direction, and similar that between the
first and the $\M$-th rows as in the $\M$ direction.
Then, our findings together with those of Ref. \cite{propp01}
lead to the following theorem applicable to all cases:

\medskip
{\it Reciprocity theorem:  Let $T({\cal M,N})$ be the number of close-packed
dimer configurations (perfect matchings) on an $\M\times\N$
simple-quartic lattice with free, periodic, or twisted boundary conditions
in either direction. (The case of
twisted boundary conditions in both directions is excluded).
If    the twisted boundary condition, if occurring, 
is in the $\M$ ($\N$) direction,    we restrict to $\M$ ($\N$) = even. 
  Then, we have

\medskip
1. $ T({\cal M,N})=\epsilon_{\cal NM}\ T(-2-{\cal M},{\cal N})$
if the boundary condition in the $\M$ direction  is free.
 
\medskip
2.  $T(\M,\N) =\epsilon_{\cal NM}\  T({-\cal M},{\cal N})$ if the boundary condition 
in the $\M$ direction 
is periodic or twisted.} 
 
%%%%%%%%%%%%%%%%%%%%%%%%%%%%%%%%%%%%%%%%%%%%%%%%%%%%%%%%
\section{Summary and Discussions}
We have evaluated the dimer generating function (\ref{gen}) for 
an $\M \times \N$ simple-quartic net embedded on a \Mo strip and a
Klein bottle for all $\M, \N$. The results are given by
(\ref{Mob-1}) - (\ref{Kln-3}).  Our results can also be written in
terms of the $q$-analogue of the Fibonacci numbers ${\cal F}_n(q)$ defined
by 
\begin{equation}
{1\over {1-qs-s^2}}  = \sum_{n=0}^\infty {\cal F}_n(q)s^n.
\end{equation}
Using the first line of (\ref{fib1}), for example,
and  the identity
\begin{eqnarray}
{\cal F}_n(q) +{\cal F}_{n-2}(q) &=& x^n + (-x)^{-n} \nonumber \\
  q & \equiv & x-x^{-1},
\end{eqnarray}
one can verify  that our results
(\ref{Mob-1}) and (\ref{Mob-3}) for the
\Mo strip are the same as those given by Tesler \cite{tesler}.
% in terms of the $q$-analogue of the Fibonacci numbers.
Details of the proof which also lead to some
new product identities involving the Fibonacci numbers
will be given elsewhere \cite{luwu02}.
% Our results for the Klein bottle are new.  
We have also deduced a reciprocity theorem for the enumeration
$T({\cal M,N})$ of dimers on an $\M\times\N$ lattice under 
arbitrary including free, periodic, and twisted
boundary conditions.

\medskip
Finally, we  point out that
the results (\ref{Mob-1}) - (\ref{Kln-3}) can be put in
 a compact expression valid for all cases as 
\be
Z_{\cal M,N}(z_h,z_v)=z_h^{{\cal MN}/2}{\rm Re}\Bigg[(1-i)
\prod_{m=1}^{[({\cal M}+1)/2]}\prod_{n=1}^{\cal N}
\Bigg(2i(-1)^{{{\cal M}\over 2}+m+1}\sin {(4n-1)\pi\over 2{\cal N}} 
+2\ X_m\Bigg)\Bigg],
\ee
where $[x]$ is the integral part of $x$, 
and 
\bea
X_m&=&\Bigg( \frac {z_v}{z_h}\Bigg)\cos \frac {m\pi} {\M+1} \hskip 1.5cm {\rm for\>\>the\>\>
Moebius \>\> strip} \nonumber \\
&=&\Bigg(\frac {z_v}{z_h}\Bigg)\sin \frac {(2m-1)\pi}{\cal M} \hskip 1cm {\rm for\>\>the\>\>Klein\>\>bottle}.
\eea
 
\section*{Acknowledgement}
Work has been supported in part by National Science Foundation Grant
DMR-9980440.

%\newpage

\end{document}